          [2001/04/25 1.1 (PWD)]
\documentclass{aa}

\usepackage{graphicx,epsfig}
\usepackage{natbib}
\bibpunct[, ]{(}{)}{;}{a}{}{,}
\begin{document}

\titlerunning{Longitudinal flux tube waves}

\title{Generation of longitudinal flux tube waves in \\
theoretical main-sequence stars:                     \\
Effects of model parameters
}

\author{
Diaa E. Fawzy \inst{1}
\and
M. Cuntz \inst{2,3}
}

\offprints{Diaa E. Fawzy}

\institute{Faculty of Engineering and Computer Sciences, Izmir University of Economics,
       35330 Izmir, Turkey; \\
       \email{diaa.gadelmavla@ieu.edu.tr}
\and
       Department of Physics, Science Hall, University of Texas at Arlington,
       Arlington, TX 76019, USA; \\
       \email{cuntz@uta.edu}
\and
       Institut f\"ur Theoretische Astrophysik, Universit\"at Heidelberg,
       69120 Heidelberg, Germany
}

\date{Received <date> / Accepted <date>}

\abstract
{Continued investigation of the linkage between magneto-acoustic energy generation in
stellar convective zones and the energy dissipation and radiative emission
in outer stellar atmospheres in stars of different activity levels.}
{We compute the wave energy fluxes carried by longitudinal tube waves along  vertically oriented
thin magnetic fluxes tubes embedded in the atmospheres of theoretical main-sequence stars based
on stellar parameters deduced by R. L. Kurucz and D. F. Gray.  Additionally, we
present a fitting formula for the wave energy flux based on the governing stellar and magnetic
parameters.}
{A modified theory of turbulence generation based on the mixing-length concept is combined with the
magneto-hydrodynamic equations to numerically account for the wave energies generated
at the base of magnetic flux tubes.}
{The results indicate a stiff dependence of the generated wave energy on the stellar and
magnetic parameters in principal agreement with previous studies.  The wave energy flux $F_{\rm LTW}$ decreases
by about a factor of 1.7 between G0~V and K0~V stars, but drops by almost two orders of magnitude
between K0~V and M0~V stars.
In addition, the values for $F_{\rm LTW}$ are significantly higher for lower in-tube magnetic field
strengths.  Both results are consistent with the findings from previous studies.}
{Our study will add to the description of magnetic energy generation in late-type main-sequence stars.
Our results will be helpful for calculating theoretical atmospheric models for stars of
different levels of magnetic activity.}


\maketitle


\section{Introduction}

An outstanding problem in stellar astrophysics concerns
the identification of physical processes responsible for the heating
of outer stellar atmospheres and the acceleration of stellar winds
(see reviews by \citealt{nara90,nara96} and \citealt{gued07}).
For the Sun and other types of stars with surface convection zones,
acoustic heating has been identified as most likely responsible for
balancing the ``basal" flux emission \citep[e.g.,][]{buch98,cunt07}.
On the other hand, it is well known that most,
if not all stars also exhibit a large amount of magnetic activity.
Thus, the chromospheres of main-sequence stars, including the Sun, are
expected to be significantly shaped by magnetically heated structure
\citep[e.g.,][]{saar94,schri96}.

There is a large body of previous work devoted to the description of
the two-component structure of stellar chromospheres.  In these models
the magnetic component of the chromosphere is typically heated by
energy dissipation of longitudinal flux tube waves.  \cite{cunt99}
computed two-component theoretical chromosphere
models for K2~V stars with different levels of magnetic activity
with the filling factor for the magnetic component determined
from an observational relationship between the measured magnetic
area coverage and the stellar rotation period.  For stars with
very slow rotation, they were able to reproduce the basal flux
limit of chromospheric emission previously identified with
non-magnetic regions.  Most notably, however, \cite{cunt99}
deduced a relationship between the Ca~II~H+K emission
and the stellar rotation rate that is consistent with the
relationship previously obtained by observations; see also
\cite{cunt98} for earlier results.

Further studies for a large spectral range of stars were given
by \cite{fawz02} based on specified values for the
magnetic filling factor.  They concluded that heating by
acoustic and longitudinal flux tube waves is able to explain
most of the observed range of chromospheric activity as gauged
by the Ca~II and Mg~II lines.  On the other hand, indirect
evidence for non-wave (i.e., reconnective) heating was also
deduced needed to explain the structure of the
highest layers of stellar chromospheres.

This type of models, as well as envisioned future models of
chromospheric heating and emission, partially motivated by
the quest of investigating the effects of UV and EUV emission
on planetary atmosphere and (potentially) the evolution of life
\citep[e.g.,][]{guin03,lamm03,gued07,cunt10}, require the
continuation of detailed simulations of magnetic wave energy
generation, including studies on longitudinal tube waves in
different types of stars, particularly main-sequence stars.
This latter goal is the focus of the present paper.

Previous work on the calculation of longitudinal tube waves
has been based on progress made by \cite{musi94} who
corrected the Lighthill-Stein theory by incorporating an
improved description of the spatial and temporal spectrum
of the turbulent convection and utilized the corrected theory
for calculating revised stellar acoustic wave energy fluxes
\citep{ulms96,ulms99}.  This type of work focused
on the generation of acoustic waves; however, it did not consider 
stellar magnetic fields.  Considering the fundamental
importance of magnetic heating in most, if not all stars,
a set of papers focused on the study of longitudinal and
transverse tube wave generation has been pursued
\citep[e.g.][]{musi89,musi95,ulms98}.  
In subsequent work, \cite{ulms01} used the
approach developed by \cite{ulms98} to compute
the wave energy fluxes carried by longitudinal tube waves
propagating along thin and vertically oriented magnetic flux
tubes that are embedded in atmospheres of late-type stars.
This numerical approach supplemented previous work by
\cite{musi00}, who analytically calculated the
longitudinal wave energy fluxes generated in
stellar convective zones.

\begin{figure}
    \begin{minipage}[t]{0.50 \textwidth}
    \includegraphics[width=0.999 \textwidth]{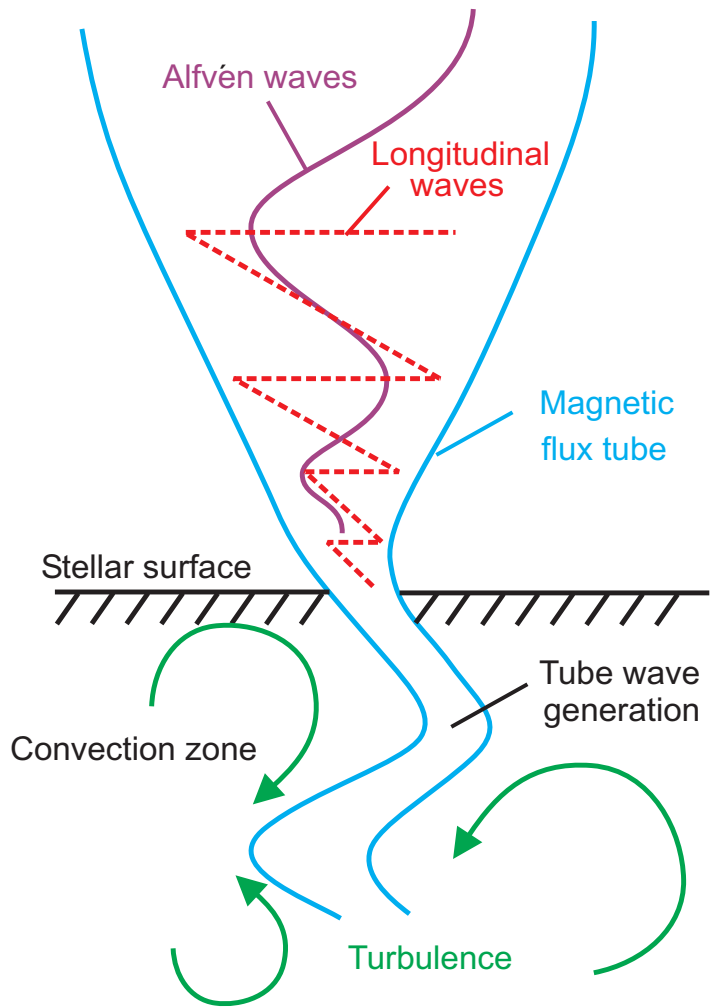}
    \end{minipage}
\caption{Diagram of a flux tube embedded into a stellar convective zone.
The squeezing point of the tube is assumed to be located at optical depth
$\tau_{5000} = 1$, coinciding with the ``stellar surface". Credit: P. Ulmschneider.
\label{fig1}}
\end{figure}

In the numerical approach by \cite{ulms98},
longitudinal tube waves are generated as a result of the
squeezing of a thin, vertically oriented magnetic 
flux tube by external pressure fluctuations produced 
by the turbulent motions in a stellar photosphere and convection 
zone, which correspond to associated velocity fluctuations.
Hence, to compute the pressure fluctuations imposed 
on the tube, it is required to know the external turbulent 
motions.  The motions are modeled by specifying the rms 
velocity amplitude and using an extended Kolmogorov turbulent
energy spectrum with a modified Gaussian frequency factor
\citep{musi94}.

The main advantage of this approach is
that it is not restricted to linear waves and that it
allows for occasionally large-amplitude waves observed
on the Sun at the photospheric level
\citep[e.g.,][]{mull85,komm91,nesi93,mull94} and also seen in detailed
time-dependent simulations of solar and stellar convection
\citep[e.g.,][]{nord90,nord91,catt91,stef93,nord97}.
Horizontal flow patterns are a notable candidate process 
for the initiation of wave modes with respect to flux tubes
(see Fig. 1); see \cite{stei09a,stei09b} for recent models of
convective flows for the Sun based on up-to-date simulations
extending toward the scale of supergranules.

The code used by \cite{ulms98}
was originally developed by \cite{herb85} who treated 
magnetic flux tubes in the so-called thin flux tube 
approximation and described them mathematically by using a set 
of one-dimensional, time-dependent and nonlinear MHD equations.
It allows to compute the instantaneous and time-averaged 
longitudinal tube wave energy fluxes and the corresponding wave 
energy spectra.  It requires specifying the strength of the 
magnetic field inside the flux tube and the height in the stellar 
atmosphere where the squeezing is assumed to take place.
The code has previously been used to calculate  
wave energy fluxes and spectra for longitudinal tube waves 
propagating in the solar atmosphere (see \citealt{fawz98}
for models of different spreading factors), and to investigate the 
dependence of these fluxes on the magnetic field strength, the rms 
velocity amplitude of turbulent motions, and the location of 
the squeezing in the atmosphere; for recent models for
other stars with non-solar metallicities see \cite{fawz10}.

The reason for reinvestigating the generation of
longitudinal flux tube waves in main-sequence stars
is three-fold.  First, we would like to use realistic
combinations of ($T_{\rm eff}$, $\log g$), with $T_{\rm eff}$
as stellar effective temperature and $\log g$ as surface
gravity, for main-sequence stars guided by up-to-date
studies by R.~L. Kurucz and D.~F. Gray.  Note that
$\log g$ is typically close to 4.5 (see Table 1).
Previous models by \cite{ulms01} and others
have been pursued for either $\log g = 4$ or $5$, thus
resulting in unnecessary interpolation errors.
Secondly, we would like to investigate the amount of
upward propagating longitudinal wave energy flux
for a wider range of stellar convective and magnetic
parameters, notably the mixing length $\alpha$ amid
recent progress made through models by \cite{stei09a,stei09b}
and others.  Thirdly, we would like to deduce a
fitting formula for the wave energy flux that
allows insight into the role of the relevant parameters
concerning that flux and, furthermore, offers a
more universal use.

Our paper is structured as follows:  In Sect. 2, we comment
on the parameters of theoretical main-sequence stars.
Additionally, we summarize the method for the computation
of longitudinal tube waves as well as the construction of
stellar flux tube models.  Our results are given in Sect. 3.
Finally, in Sect. 4 we present the summary and conclusions.

\begin{table}
\caption{Theoretical main-sequence stars}
\centering
\vspace{0.05in}
\vspace{0.05in}
\begin{tabular}{l c c c c}
\hline
\hline
\noalign{\vspace{0.03in}}
Sp. Type & $T_{\rm eff}$ & $R$         & $M$         & $\log~g$ \\
...      & (K)           & ($R_\odot$) & ($M_\odot$) & ...      \\
\noalign{\vspace{0.03in}}
\hline
\noalign{\vspace{0.03in}}
 F0~V  &  7178 &    1.620  &   1.600  &   4.223 \\
 F1~V  &  7042 &    1.541  &   1.560  &   4.255 \\
 F2~V  &  6909 &    1.480  &   1.520  &   4.279 \\   
 F3~V  &  6780 &    1.453  &   1.480  &   4.283 \\   
 F4~V  &  6653 &    1.427  &   1.440  &   4.287 \\    
 F5~V  &  6528 &    1.400  &   1.400  &   4.292 \\    
 F6~V  &  6403 &    1.333  &   1.330  &   4.312 \\    
 F7~V  &  6280 &    1.267  &   1.260  &   4.333 \\    
 F8~V  &  6160 &    1.200  &   1.190  &   4.355 \\    
 F9~V  &  6047 &    1.155  &   1.120  &   4.362 \\    
 G0~V  &  5943 &    1.120  &   1.050  &   4.360 \\    
 G1~V  &  5872 &    1.100  &   1.022  &   4.364 \\    
 G2~V  &  5811 &    1.080  &   0.994  &   4.368 \\   
 G3~V  &  5760 &    1.037  &   0.967  &   4.392 \\   
 G4~V  &  5708 &    0.993  &   0.940  &   4.417 \\   
 G5~V  &  5657 &    0.950  &   0.914  &   4.443 \\   
 G6~V  &  5603 &    0.937  &   0.888  &   4.443 \\   
 G7~V  &  5546 &    0.923  &   0.863  &   4.443 \\   
 G8~V  &  5486 &    0.910  &   0.838  &   4.443 \\   
 G9~V  &  5388 &    0.870  &   0.814  &   4.469 \\   
 K0~V  &  5282 &    0.830  &   0.790  &   4.497 \\   
 K1~V  &  5169 &    0.790  &   0.766  &   4.527 \\   
 K2~V  &  5055 &    0.750  &   0.742  &   4.558 \\   
 K3~V  &  4973 &    0.730  &   0.718  &   4.567 \\   
 K4~V  &  4730 &    0.685  &   0.694  &   4.608 \\   
 K5~V  &  4487 &    0.640  &   0.670  &   4.651 \\   
 K6~V  &  4294 &    0.601  &   0.643  &   4.689 \\   
 K7~V  &  4133 &    0.565  &   0.614  &   4.722 \\   
 K8~V  &  4006 &    0.533  &   0.582  &   4.749 \\   
 K9~V  &  3911 &    0.505  &   0.547  &   4.770 \\   
 M0~V  &  3850 &    0.480  &   0.510  &   4.783 \\    
\noalign{\vspace{0.03in}}
\hline
\end{tabular}
\end{table}

\section{Methods}

\subsection{Comments on the theoretical main-sequence stars}

Stellar parameters for theoretical main-sequence stars have been
deduced by \cite{gray05}; see his Table B.1.  His values,
notably $T_{\rm eff}$ and $\log g$, serve as basis for the
present study.  We also improved the accuracy of the $\log g$
values if more accurate values for the stellar masses and
stellar radii were given.  For stellar spectral types with
no data given, we calculated those data using biparabolic
interpolation.  The stellar data are summarized in Table~1.

Another set of spectral models has been constructed by
R.~L.~Kurucz and collaborators.  They take into account millions
or hundred of millions of lines for a large array of atoms and molecules;
see, e.g., \cite{cast04} and \cite{kuru05} for technical details.
These models indicate very similar effective temperatures compared
to the models by \cite{gray05} for most types of stars.  However,
stellar spectral types of K5~V and below, the indicated effective
temperatures of R.~L.~Kurucz are consistently lower noting that
the difference amounts to nearly 300~K for spectral type M0~V.
Therefore, we assumed average values between the models by
D.~F. Gray and R.~L.~Kurucz for stars of spectral spectral type
K5~V and M0~V in the following.

\begin{figure}
    \begin{minipage}[t]{0.50 \textwidth}
    \includegraphics[width=0.999 \textwidth]{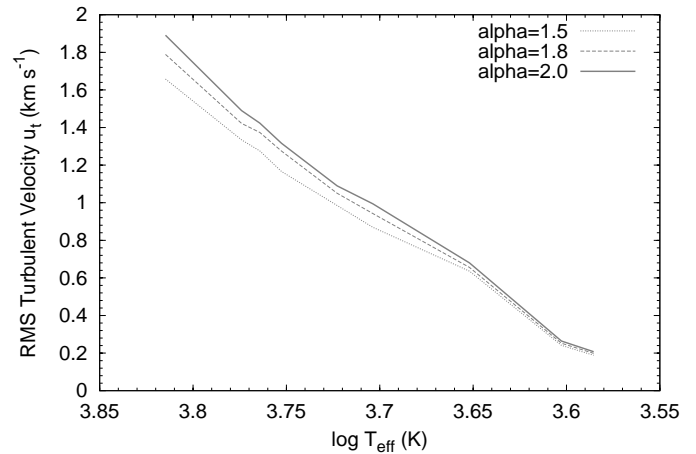}
    \end{minipage}
\caption{Root mean square turbulent velocities at the squeezing point ($\tau_{5000}=1$)
for the set of theoretical main-sequence stars for different values of the
mixing-length parameter $\alpha$.
\label{fig2}}
\end{figure}

\begin{figure}
    \begin{minipage}[t]{0.50 \textwidth}
    \includegraphics[width=0.999 \textwidth]{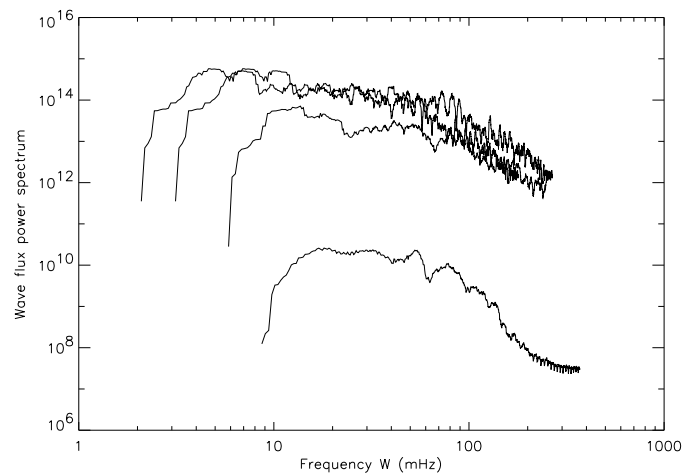}
    \end{minipage}
\caption{
Power spectra of the instantaneous wave energy flux in
flux tubes with $B/B_{\rm eq} = 0.85$ for
F5~V, G5~V, K5~V, and M0~V stars (from top to bottom)
as a function of circular frequency $\omega$.  The
mixing-length parameter is assumed as $\alpha = 2$.
\label{fig3}}
\end{figure}

\begin{table}
\caption{Wave energy flux$^a$ for different parameters $\alpha$ and $\eta$}
\centering
\vspace{0.05in}
\vspace{0.05in}
\begin{tabular}{l c c c c}
\hline
\hline
\noalign{\vspace{0.03in}}
Sp.Type & $T_{\rm eff}$ & $F_{\rm LTW}$ & $F_{\rm LTW}$ & $F_{\rm LTW}$ \\
...     & (K)           & ...           & ...           & ...           \\
\noalign{\vspace{0.03in}}
\hline
\noalign{\vspace{0.03in}}
... & ... & $\eta = 0.75$ & $\eta = 0.85$ & $\eta = 0.95$ \\
\noalign{\vspace{0.03in}}
\hline
\noalign{\vspace{0.03in}}
\multicolumn{2}{c}{$\alpha = 1.5$} & & \\
\noalign{\vspace{0.03in}}
\hline
\noalign{\vspace{0.03in}}
F5~V	&  6528  &	6.03E8  &  4.05E8  &  9.41E7  \\
F8~V  &  6160  &  5.59E8  &  3.10E8  &  9.14E7  \\
G0~V	&  5943  &	4.75E8  &  2.94E8  &  7.91E7  \\
G2~V	&  5811  &	4.50E8  &  2.62E8  &  7.91E7  \\
G5~V	&  5657  &	3.56E8  &  2.20E8  &  6.53E7  \\
G8~V	&  5486  &  3.27E8  &  1.98E8  &  5.85E7  \\
K0~V  &  5282  &	2.54E8  &  1.56E8  &  4.77E7  \\
K2~V	&  5055  &	1.92E8  &  1.10E8  &  3.42E7  \\
K5~V	&  4487  &	1.14E8  &  6.83E7  &  2.12E7  \\
K8~V	&  4006  &	4.47E6  &  2.87E6  &  9.82E5  \\
M0~V	&  3850  &	2.26E6  &  1.33E6  &  5.36E5  \\
\noalign{\vspace{0.03in}}
\hline
\noalign{\vspace{0.03in}}
\multicolumn{2}{c}{$\alpha = 1.8$} & & \\
\noalign{\vspace{0.03in}}
\hline
\noalign{\vspace{0.03in}}
F5~V	&  6528  &	8.99E8  & 5.00E8   &  1.25E8  \\
F8~V  &  6160  &  7.51E8  & 4.62E8   &  1.15E8  \\
G0~V	&  5943  &	5.69E8  & 3.40E8   &  9.73E7  \\
G2~V	&  5811  &	5.63E8  & 3.02E8   &  9.05E7  \\
G5~V	&  5657  &	5.29E8  & 2.78E8   &  7.96E7  \\
G8~V	&  5486  &  3.98E8  & 2.37E8   &  7.32E7  \\
K0~V  &  5282  &	2.93E8  & 1.82E8   &  5.60E7  \\
K2~V	&  5055  &	2.69E8  & 1.63E8   &  4.72E7  \\
K5~V	&  4487  &	1.30E8  & 8.15E7   &  2.50E7  \\
K8~V	&  4006  &	5.36E6  & 3.50E6   &  1.24E6  \\
M0~V	&  3850  &	2.28E6  & 1.44E6   &  6.49E5  \\
\noalign{\vspace{0.03in}}
\hline
\noalign{\vspace{0.03in}}
\multicolumn{2}{c}{$\alpha = 2.0$} & & \\
\noalign{\vspace{0.03in}}
\hline
\noalign{\vspace{0.03in}}
F5~V	&  6528  &	1.12E9  &  6.07E8  &  1.47E8  \\
F8~V  &  6160  &	8.51E8  &  4.92E8  &  1.34E8  \\
G0~V	&  5943  &	6.77E8  &  3.96E8  &  1.11E8  \\
G2~V	&  5811  &	6.35E8  &  3.65E8  &  1.06E8  \\
G5~V	&  5657  &	5.81E8  &  3.35E8  &  1.01E8  \\
G8~V	&  5486  &	4.79E8  &  2.86E8  &  8.05E7  \\
K0~V  &  5282  &	3.39E8  &  1.95E8  &  6.39E7  \\
K2~V	&  5055  &	2.98E8  &  1.86E8  &  5.81E7  \\
K5~V	&  4487  &	1.44E8  &  9.19E7  &  2.86E7  \\
K8~V	&  4006  &	6.79E6  &  4.37E6  &  1.57E6  \\
M0~V	&  3850  &	3.32E6  &  2.11E6  &  7.62E5  \\
\noalign{\vspace{0.03in}}
\hline
\end{tabular}
\vspace{0.05in}
\begin{list}{}{}
\item[]$^a$The unit of $F_{\rm LTW}$ is erg~cm$^{-2}$~s$^{-1}$.
\end{list}
\end{table}

\subsection{Convective zone models and turbulent velocities}

The method for calculating wave energy fluxes carried by 
longitudinal tube waves adopted in the present paper has been
described in detail by \cite{ulms01}.
Thus, it is not necessary to present an intricate discussion in
the following.  In the solar application, it is possible to
select many model parameters and characteristic values directly
from observations.  However, for stars other than the Sun such
data are mostly unavailable.   Therefore, we need to discuss in
some detail the physical reasoning behind our choice of relevant
parameters used in our calculations.

In the current approach, the magnetic flux tubes are embedded
in nonmagnetized photospheric convection zones (see Fig.~1).
Considering that the
interaction between the flux tubes and the convective
turbulence is the driving mechanism for the generation of
longitudinal tube waves, among other waves, models of the
stellar convection zones are required.  Guided by previous
studies, it is assumed that the squeezing of the tube
is symmetric with respect to the tube axis.  
The computed pressure fluctuations are subsequently
translated into gas pressure and magnetic field fluctuations
inside the tube assuming horizontal pressure balance.
Finally, the internal velocity perturbation resulting
from the internal pressure fluctuation is calculated. 
This internal velocity served as a boundary condition 
in the numerical simulation of the generation of the
longitudinal tube waves. 

Both numerical simulations of stellar convection and mixing length 
models show that the maximum convective velocities occur at optical 
depths of $\tau_{5000}\approx 10$ to 100.  For example, \cite{stef93}
found in his time-dependent solar numerical convection calculations 
that maximum convective velocities $v_{\rm CMax} \simeq 2.8$ km~s$^{-1}$
are reached at $\tau_{5000} \approx 50$ and that these values can be 
reproduced via mixing length theory with a mixing length parameter 
of $\alpha \simeq 2$.  The value $\alpha = 2$ is furthermore
indicated by time-dependent hydrodynamic simulations of stellar convection
for stars other than the Sun \citep{tram97} as well as by a 
careful fitting of evolutionary tracks of the Sun with its present 
luminosity, effective temperature and age \citep{schro96}.  

Nevertheless, there is still some debate about the most appropriate
value of $\alpha$.  \cite{nord90} originally pursued
detailed numerical simulations based on 3-D hydrodynamics coupled with
3-D non-grey radiative transfer for stars similar to Procyon (F5~IV-V),
$\alpha$ Cen~A (G2~V), $\beta$~Hyi (G2~IV), and $\alpha$ Cen~B (K1~V).
They concluded a mixing-length parameter of $\alpha = 1.5$ (or
slightly higher), although the mixing length concept appeared to be
problematic at photospheric heights.  A mixing length of 1.5 was also
used by \cite{cunt99} in their two-component theoretical chromosphere
models for K2~V stars with different levels of magnetic activity.
Even though the deduced relationship between the Ca~II~H+K emission
and the stellar rotation rate was found to be largely consistent with the
observed relationship, the agreement could probably be improved if
a somewhat higher longitudinal wave energy flux, corresponding to a
slightly larger mixing length parameter, was adopted.  Recently,
\cite{stei09a,stei09b} pursued updated state-of-the-art simulations
of solar convection zone extending toward the scale of supergranules
indicating a mixing length parameter of $\alpha \simeq 1.8$.
For these reasons, we will calculate a set of models concerning
wave energy generation of longitudinal tube wave for a set of
$\alpha$ values, which are $\alpha$ = 1.5, 1.8, and 2.0.

\subsection{Computation of stellar magnetic flux tube models}

Our treatment of stellar convection associated with the facilitation
of stellar flux tube models is akin to that described by
\cite{ulms96}.  In this approach, information is needed
about velocities of turbulent motions in the overshooting layer near
the stellar surface, where the squeezing of the magnetic flux tube
is assumed to occur.  
Steffen's numerical calculations show that the rms velocities decrease 
toward the solar surface and reach a plateau in the overshooting layer.  
Between $\tau_{5000}=1$ and $10^{-4}$, he finds values of $v_{\rm rms} = 
1.4$ km~s$^{-1}$, which are essentially independent of height.
\cite{ulms98} adopted for the Sun a variety of observed rms velocity 
amplitudes $u_t$ in the range $0.9<u_t<1.9$ km~s$^{-1}$, and showed the 
dependence of the computed fluxes on this velocity.  
For stars, these velocities cannot be determined from observations; thus,
we follow \cite{ulms01} by assuming that the rms velocity
fluctuations at the squeezing points are given by $u_t = v_{\rm CMax}/2$.
The values of $v_{\rm t}$ and
$v_{\rm CMax}$ (see Fig.~2) are evaluated from stellar convection zone
models based on the adopted mixing length parameter.
The numerical factor 2 used in our calculations ensures that
in our approach the considered convective velocities are always
lower than the local speed of sound.

After specifying the rms velocities that are responsible for the 
wave generation, we also determine the height in stellar 
atmospheres, where the most efficient squeezing of magnetic flux tubes 
takes place.  We are guided by earlier studies performed by
\cite{ulms98}, who pointed out that shifting the height of the
excitation point did not change much the resulting wave 
energy fluxes for the Sun.  Based on these results, we take the 
squeezing point to be located at optical depth $\tau_{5000} = 1$ for 
all considered stars.  This depth is commonly taken as the zero 
height level in stellar atmosphere computations.


\begin{table*}
\caption{Wave energy flux$^a$ for different values of $\eta$ based on $\alpha = 2$} 
\centering
\vspace{0.05in}
\vspace{0.05in}
\begin{tabular}{l c c c c c c c c c c}
\hline
\hline
\noalign{\vspace{0.03in}}
Sp.Type & $T_{\rm eff}$ & $F$(tot) & $F$(prop) & $F$(prop+up) & $F$(tot) & $F$(prop) & $F$(prop+up) & $F$(tot) & $F$(prop) & $F$(prop+up) \\
...     & (K)           & ...      & ...       & ...          & ...      & ...       & ...          & ...      & ...       & ...          \\
\noalign{\vspace{0.03in}}
\hline
\noalign{\vspace{0.03in}}
... & ... & \multicolumn{3}{c}{$\eta = 0.75$} & \multicolumn{3}{c}{$\eta = 0.85$} & \multicolumn{3}{c}{$\eta = 0.95$} \\
\noalign{\vspace{0.03in}}
\hline
\noalign{\vspace{0.03in}}
F2~V	&  6909  &   ...   &   ...   &   ...   &  5.18E8 &  4.37E8 &	6.26E8 & 1.17E8 &	1.01E8 &	1.41E8 \\
F5~V	&  6528  &	8.23E8 &  6.97E8 &  1.12E9 &  5.04E8 &  4.13E8 &	6.07E8 & 1.27E8 &	1.08E8 &	1.47E8 \\
F8~V  &  6160  &	6.76E8 &  6.02E8 &  8.51E8 &  4.10E8 &  3.59E8 &	4.92E8 & 1.09E8 &	9.48E7 &	1.34E8 \\
G0~V	&  5943  &	5.45E8 &  4.71E8 &  6.77E8 &  3.22E8 &  2.76E8 &	3.96E8 & 9.75E7 &	8.11E7 &	1.11E8 \\
G2~V	&  5811  &	4.91E8 &  4.26E8 &  6.35E8 &  2.96E8 &  2.53E8 &	3.65E8 & 8.70E7 &	7.56E7 &	1.06E8 \\
G5~V	&  5657  &	4.75E8 &  3.93E8 &  5.81E8 &  2.73E8 &  2.32E8 &	3.35E8 & 8.45E7 &	7.22E7 &	1.01E8 \\
G8~V	&  5486  &	3.86E8 &  3.29E8 &  4.79E8 &  2.38E8 &  2.02E8 &	2.86E8 & 6.76E7 &	5.87E7 &	8.05E7 \\
K0~V  &  5282  &	2.95E8 &  2.30E8 &  3.39E8 &  1.66E8 &  1.39E8 &	1.95E8 & 5.47E7 &	4.68E7 &	6.39E7 \\
K2~V	&  5055  &	2.52E8 &  2.08E8 &  2.98E8 &  1.55E8 &  1.26E8 &	1.86E8 & 4.86E7 &	3.90E7 &	5.81E7 \\
K5~V	&  4487  &	1.29E8 &  8.93E7 &  1.44E8 &  8.08E7 &  5.93E7 &	9.19E7 & 2.53E7 &	1.83E7 &	2.86E7 \\
K8~V	&  4006  &	5.55E6 &  3.36E6 &  6.79E6 &  3.74E6 &  2.14E6 &	4.37E6 & 1.25E6 &	6.95E5 &	1.57E6 \\
M0~V	&  3850  &	2.29E6 &  1.36E6 &  3.32E6 &  1.57E6 &  8.23E5 &	2.11E6 & 5.37E5 &	2.86E5 &	7.62E5 \\
\noalign{\vspace{0.03in}}
\hline
\end{tabular}
\vspace{0.05in}
\begin{list}{}{}
\item[]$^a$The unit of $F$(tot), $F$(prop), and $F$(prop+up) is erg~cm$^{-2}$~s$^{-1}$.  Note that 
$F$(prop+up) $\equiv$ $F_{\rm LTW}$.
\end{list}
\end{table*}

For the computation of stellar magnetic flux tubes we consider
the ``thin flux tube approximation" \citep[e.g.,][]{spru81}.
Two dimensional modeling of magnetic flux tubes by \cite{hasa03}
indicates that this approach renders reasonable results
if the models do not extend beyond a small number of scale heights above
the stellar surface; note that this approximation is fully compatible
with our study.  In the following,
we augment the commonly used concept of solar magnetic flux tubes
\citep[e.g.,][]{sten78,sola93} to stellar flux tubes; see
\cite{sola96} for further discussion.  Stellar magnetic flux tubes
at the stellar surface are assumed to have diameters of roughly equal
to the local pressure scale height like in case of the Sun.

For the Sun, the magnetic flux tubes within the solar photosphere
have field strengths on the order $B_0 = 1500$~G \citep[e.g.,][]{sola93}.
Taking $p_e=1.17\cdot 10^5\ \rm dyn\ cm^{-2}$ from model C of
\cite{vern81} at the height $z=0$ where $\tau_{5000}=1$ yields an
equipartition field strength ($B_{\rm eq}^2/8\pi = p_e$) of $B_{\rm eq}=1716$~G.
This corresponds to a ratio $B/B_{\rm eq}=0.875$, which may or may not be
typical for stellar flux tubes.  Since this ratio is likely
to vary even on the Sun \citep[e.g.,][]{schri00}, it is appropriate
to consider a range of that ratio for the sake of comprehensiveness.
Therefore, we take $B/B_{\rm eq} = 0.75$, 0.85, and 0.95 for
our stellar flux tube models.  This allows us to deduce an appropriate
set of energy fluxes for upward propagating longitudinal tube waves for
specified values of $\alpha = 1.5$, 1.8, and 2.0, resulting in
a total of 9 models per theoretical target star.

\begin{figure}
    \begin{minipage}[t]{0.50 \textwidth}
    \includegraphics[width=0.999 \textwidth]{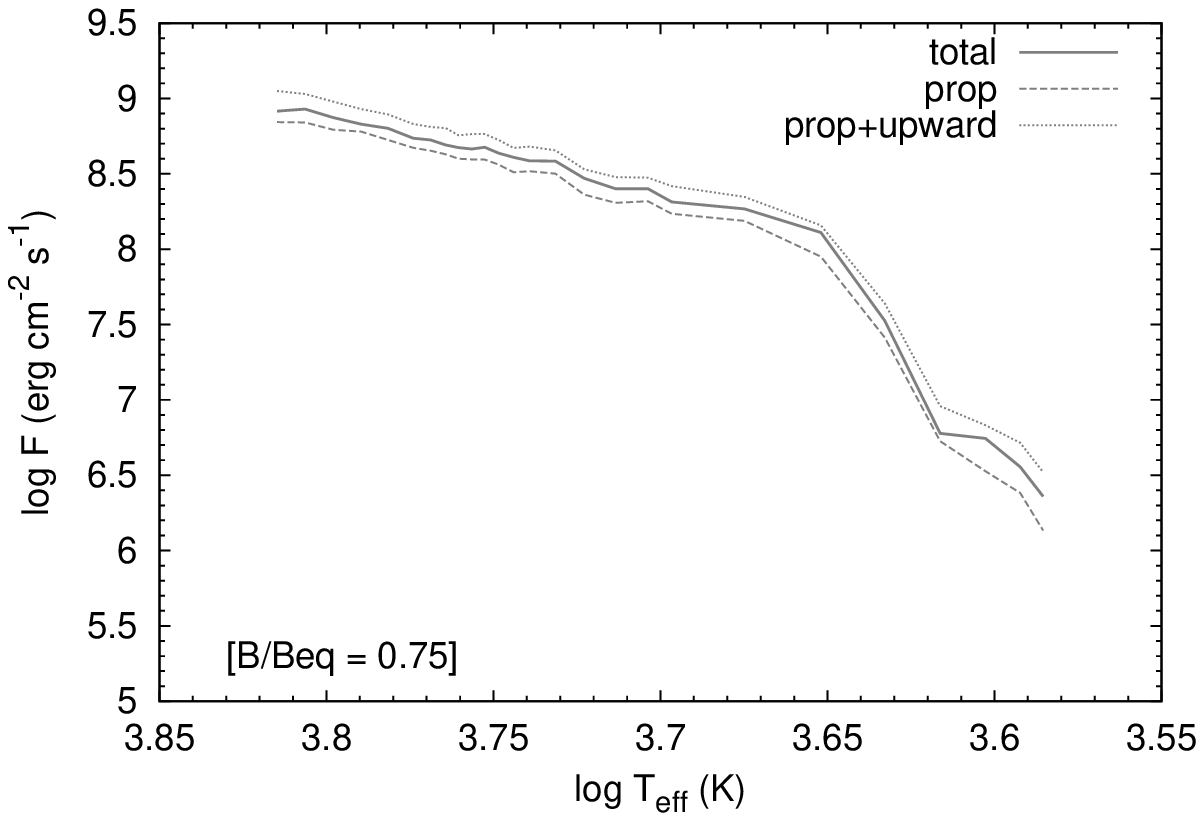}
    \includegraphics[width=0.999 \textwidth]{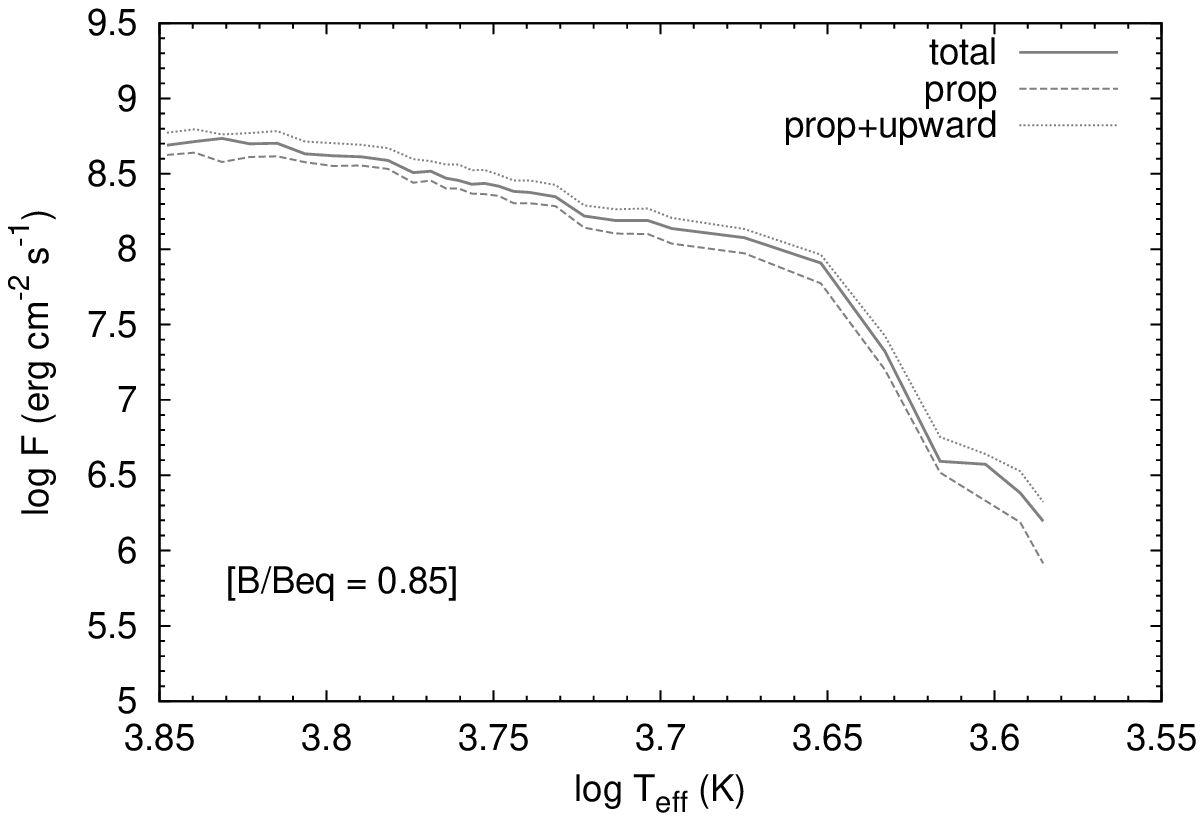}
    \includegraphics[width=0.999 \textwidth]{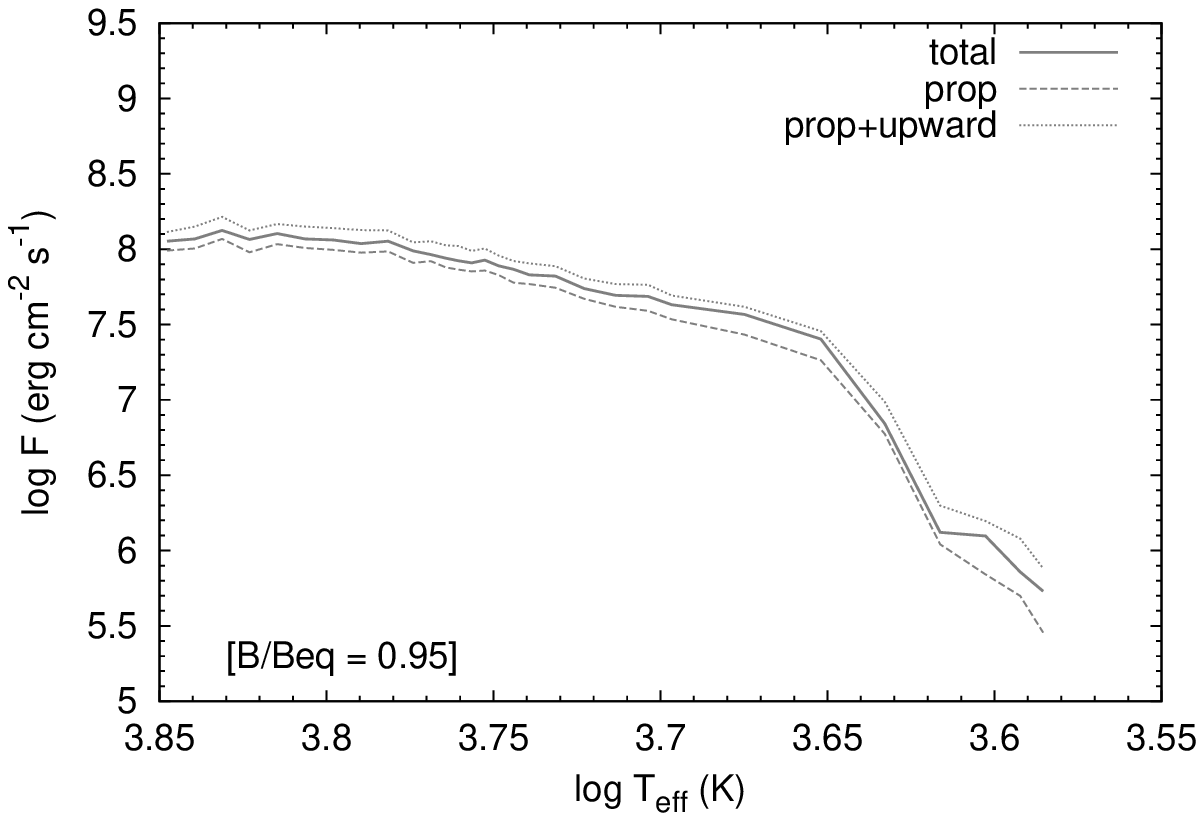}
    \end{minipage}
\caption{
Different types of energy fluxes (see text) for the set of theoretical main-sequence
stars for $\alpha = 2$.  Results are given for $B/B_{\rm eq} = 0.75$ (top),
0.85 (middle), and 0.95 (bottom).
\label{fig4}}
\end{figure}

\begin{figure}
    \begin{minipage}[t]{0.50 \textwidth}
    \includegraphics[width=0.999 \textwidth]{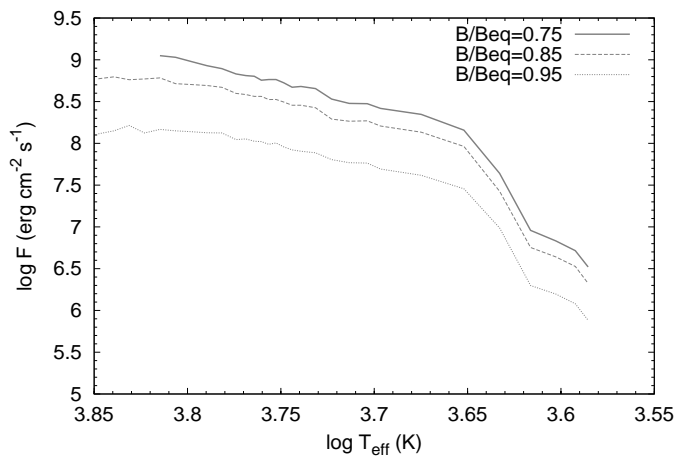}
    \end{minipage}
\caption{
Upward propagating energy fluxes, also referred to as $F_{\rm LTW}$,
for the set of theoretical main-sequence stars for $\alpha = 2$ and
different values of $\eta = B/B_{\rm eq}$.
\label{fig5}}
\end{figure}

\section{Results and discussion}

\subsection{Computation of wave energy fluxes}

The instantaneous and time-averaged tube wave energy fluxes are computed
by employing a modified time-dependent wave code based on an earlier version
by \cite{herb85}.  The external turbulent motion is translated into
internal pressure and velocity fluctuations.  The time-averaged wave energy
fluxes at the squeezing point are computed for each combination of
effective temperature $T_{\rm eff}$, surface gravity $\log g$,
$\eta = B/B_{\rm eq}$ and mixing-length parameter $\alpha$ for flux tubes
embedded in the atmospheres of the main-sequence stars as considered.

For the evaluation of reliable time-averaged wave energy fluxes and due to the
spiky nature of the instantaneous wave energy fluxes, the wave propagation code
has to be run over a time of about $35 P_{D}$ with $P_{D}$ as the wave period
of the Defouw cut-off frequency \citep{defo76}.  The generated waves include
both propagating and non-propagating wave energy fluxes.  We apply a high pass
filter at the Defouw cut-off frequency, $\omega_{D}$, on the velocity and pressure
fluctuations inside the flux tubes to separate the propagating waves and 
to compute the time-averaged upward propagating wave components.
The position of the filter is indicated by the low frequency cut-off depicted
in Fig.~3, which features a comparison of the
power spectra for different types of stars.  Note that the power spectra
reach their maxima at roughly $2 \omega_{D}$ and decrease toward
higher frequencies.

Figure~4 and 5 as well as Table~2 and 3 show the computed wave energy
of the magnetic flux tube. Actually, Table 3 for the case of $\alpha = 2$
also distinguishes between the total energy flux, propagating energy flux
and upward propagating energy flux; see \cite{ulms01} for detailed definitions.
The wave energy flux $F_{\rm LTW}$ identified as upward propagating energy
flux shows a characteristic behaviour as function of the model parameters,
notably the stellar spectral type (or $T_{\rm eff}$).  It is found that
the value of $F_{\rm LTW}$ decreases by about a factor of 1.7 between G0~V
and K0~V stars, and drops by almost two orders of magnitude between K0~V
and M0~V stars.  This is a direct consequence of the action of stellar
convection, noting that in relatively hot main-sequence stars there is an
enhanced efficiency in the creation of turbulence, reflected in higher
values of the rms velocity $u_t$ (see Fig.~2).

Additionally, we also found that the wave energy flux $F_{\rm LTW}$ shows
significant dependence on the magnetic field strengths inside of the
tubes, given as $\eta = B/B_{\rm eq}$ as well as some minor dependence
on the mixing-length parameter $\alpha$.  In fact, the wave energy
fluxes are found to increase with decreasing magnetic field strengths
inside the flux tube, a phenomenon that can be explained by the decreasing
stiffness of the magnetic tube; see Fig.~5 for details.  This behaviour
was previously also pointed out by \cite{ulms01}.

Moreover, it is found that the wave energy flux for upward propagating waves
is also somewhat sensitive to the choice of $\alpha$ concerning the stellar
convective zone.  Note that an increase of the mixing length from $\alpha$ = 1.5
to 2.0 amplifies the simulated longitudinal wave energy fluxes by a factor of
approximately 1.45, which is most notable for relatively hot main-sequence stars.
This can be explained by the fact that for higher values of $\alpha$, the convective
zones are more efficient concerning the generation of turbulence, thus resulting
in thus higher values of the rms velocity $u_t$ (see Fig.~2).

\subsection{Derivation of fitting formulas}

The amount of data for $F_{\rm LTW}$ for stars of different spectral types 
and the various choices for $\eta$ and the mixing-length parameter $\alpha$
are a strong motivation for the derivation of a fitting formulas.  Our main focus
are stars between $T_{\rm eff} = 6000$~K (i.e., spectral type F9.5~V) and
stars of spectral-type mid-K.  Stars below mid-K are increasingly dominated
by flare activity or other processes as, e.g., magnetic reconnection
\citep[e.g.,][]{nara96}, which makes the existence of an accurate
fitting formula less urgent.  Nevertheless, we extended our
fitting formula toward stars of spectral type M0~V (with $T_{\rm eff} =
3850$~K) while reproducing the wave energy flux of M0~V stars with a
precision of 10$^{-3}$.

However, no attempt has been made to accurately
reproduce the knee in the wave energy flux function near F8~V as this
would have led to an additional complication in our fitting formula without
obvious merits.  This approach is motivated by the finding that energy
dissipation by longitudinal tube waves appears to be less important in
dwarfs stars of spectal type mid-K to M compared to, e.g., flare heating
as pointed out by \cite{fawz02} in their comparison between empirical
and theoretical radiative chromospheric emission losses.

Our formula is given as follows:
\begin{equation}
F_{\rm LTW} \ = \ 2.55 \cdot 10^8 \
       T_{\rm rel}^\epsilon \ e^{{\mu} T_{\rm dif}} \ Z(\alpha,\eta)  \ \
       {\rm erg~cm}^{-2}~{\rm s}^{-1} 
\end{equation}
\noindent
with $T_{\rm rel} = T_{\rm eff} / T_{{\rm eff},\odot}$ noting that
$T_{{\rm eff},\odot} = 5777$~K.  In addition, $Z(\alpha,\eta)$ is
given as
\begin{equation}
Z(\alpha,\eta) \ = \ \alpha^{1.25} \cdot \Bigl({\eta \over 0.75}\Bigr)^\gamma
\end{equation}
with $\eta = B/B_{\rm eq}$ and $\gamma = - 4.5 - 30 \cdot \vert \eta - 0.85 \vert$.
Furthermore, $T_{\rm dif}$ is defined as Min$(0,T_{\rm eff} - 4500$~K),
which means that $T_{\rm dif}$ is zero for $T_{\rm eff} \ge 4500~{\rm K}$
and takes a negative value otherwise.  Note that the parameters
$\epsilon$ and $\mu$ of Eq.~(1) weakly depend on $\alpha$ and $\eta$;
see Table~4 and 5 for detailed information.  If a reduced level of
accuracy is permitted, it might be appropriate to use average values for
$\alpha$ and $\mu$, given as $\alpha = 5.21$ and $\mu = 4.66 \cdot 10^{-3}$,
respectively.

Our formula for the upward propagating longitudinal wave energy flux has
been subjected to thorough testing for stars of spectral type F9.5~V, G2~V,
G5~V, G8~V, K0~V, K2~V, and K5~V.  The simulated data for the F9.5~V star
were obtained via logarithmic interpolation between the data for the F8~V
and G0~V stars.  Detailed information on the tests is given in Appendix A.

\begin{table}
\caption{Data for the parameter $\epsilon$}
\centering
\vspace{0.05in}
\vspace{0.05in}
\begin{tabular}{l c c c}
\hline
\hline
\noalign{\vspace{0.03in}}
$\alpha$ & $\eta = 0.75$ & $\eta = 0.85$ & $\eta = 0.95$ \\
\noalign{\vspace{0.03in}}
\hline
\noalign{\vspace{0.03in}}
 1.5  & 5.24 & 5.30 & 4.95  \\
 1.8  & 5.51 & 5.11 & 4.87  \\
 2.0  & 5.67 & 5.31 & 4.90  \\
\noalign{\vspace{0.03in}}
\hline
\end{tabular}
\end{table}

\begin{table}
\caption{Data for the parameter $\mu$}
\centering
\vspace{0.05in}
\vspace{0.05in}
\begin{tabular}{l c c c}
\hline
\hline
\noalign{\vspace{0.03in}}
$\alpha$ & $\eta = 0.75$ & $\eta = 0.85$ & $\eta = 0.95$ \\
\noalign{\vspace{0.03in}}
\hline
\noalign{\vspace{0.03in}}
 1.5  &  4.78E-3 &  4.69E-3 &  4.45E-3 \\
 1.8  &  4.95E-3 &  5.04E-3 &  4.55E-3 \\
 2.0  &  4.47E-3 &  4.53E-3 &  4.49E-3 \\
\noalign{\vspace{0.03in}}
\hline
\end{tabular}
\end{table}

\section{Summary and conclusions}

We studied the generation of longitudinal waves in stellar magnetic flux
tubes of theoretical main-sequence stars.  Our results are commensurate
with those obtained from previous studies, especially the work by
\cite{ulms01}.  Our investigations show that through
nonlinear time-dependent responses of stellar magnetic flux tubes to
continuous and impulsive external turbulent pressure fluctuations,
longitudinal tube waves were effectively produced via dipole emission. 
Furthermore, the shapes of the computed power spectra were found
to be similar for stars of different effective temperature.  Moreover,
the longitudinal wave energy fluxes are found to increase with
higher effective temperature, i.e., stars of earlier spectral types.

As part of our study, we investigated the role of the magnetic field
strength inside of the tube $B$ as well as that of the adopted
convective model characterized by the mixing-length parameter $\alpha$
concerning the generated wave energy flux.  We found that the
computed wave energy flux strongly depends on the strength of the
magnetic field as already discussed in previous studies
\citep[e.g.,][]{ulms98,ulms01}.
For a given spectral type, the flux is considerably higher in tubes
with a field strength of $B/B_{\rm eq} = 0.75$ compared
to $B/B_{\rm eq} = 0.95$, although the difference as a function of
spectral type is not as large as previously pointed out by
\cite{ulms01} owing to the differences in stellar surface gravity
for the different types of stars.  Note that the difference concerning
the magnetic field strength of $B/B_{\rm eq} = 0.75$ and 0.95 is found
to be a factor of 7.6, 6.1, 5.3, and 4.4 (for $\alpha = 2$) for stars
of spectral type F5~V, G0~V, K0~V, and M0~V, respectively.  This
difference exhibits a noticeable, albeit little dependence on the
mixing-length parameter $\alpha$ for stars hotter than G2~V,
rooted in the behaviour of the adopted root mean square velocity at
the squeezing point of the tube.

Another aspect of our study was to consider a limited range of
the mixing-length parameter $\alpha$, which are 1.5, 1.8, and 2.0.
It was found that an increase of the mixing length from $\alpha$ = 1.5
to 2.0 enhances the computed energy fluxes by a factor of about 1.45,
corresponding to a proportionality of $\alpha^{1.25}$.
This relationship can be compared with previous findings for
acoustic energy generation, which show a dependence such as
$\alpha^{2.8}$ \citep{bohn84} or $\alpha^{3.8}$ in updated models
by \cite{musi94}.  The relatively weak influence of
$\alpha$ on the amount of upward propagating wave energy flux
$F_{\rm LTW}$ is apparently due to the fact that the latter is
largely controlled by magnetic processes, as also evidenced by
the strong influence of $\eta$ up to $\eta^{-7.5}$,
rather than controlled by convective processes, even though
the latter are key for the excitement of the tubes subsequently
resulting in magnetic wave generation.

Note that the strong dependence of the generated wave energy fluxes
on the stellar and magnetic parameters is in general agreement
with the findings from previous studies, although some noticeable
differences are attained.  The wave energy flux $F_{\rm LTW}$
decreases by about a factor of 1.7 between the G0~V and K0~V stars, and
drops by almost two orders of magnitude between the K0~V and M0~V stars.
For $\alpha = 2$ at a fixed value of $T_{\rm eff} = 5000$~K,
\cite{ulms01} deduced a fitting formula for the behaviour
of $F_{\rm LTW}$ as function of $B/B_{\rm eq}$, which is found to be
commensurate with the results obtained in our current study.  On the
other hand, the fitting formula for the wave energy flux given in our
paper is more general than any of the previously deduced formulas
as it is applicable to a large range of stellar effective temperatures
and furthermore allows insight into the role of the governing magnetic
and convective parameters concerning the amount of generated upward
propagating wave energy flux.  Therefore, it is of interest to future solar
and stellar physics studies as it allows flexibility both concerning the
mixing-length parameter $\alpha$ and the magnetic parameter
$\eta = B/B_{\rm eq}$.  Note that even for the Sun, $\eta$ is expected
to exhibit considerable spatial and temporal fluctuations across the
surface as implied by previous studies of solar physics research; see,
e.g., \cite{schri00} for background information.

\bigskip

\appendix

\section{tests for the fitting formula}

For testing our fitting formula (see Eq.~1) we considered two different metrices,
i.e., the linear and the quadratic metric, which allow us to assess the
accuracy of the formula.  The linear metric is given as
\begin{equation}
{\rm dev}_1 \ = \  {1 \over N} \sum_{i=1}^N \vert F_{\rm LTW}^{\ast} - F_{\rm LTW} \vert \ ,
\end{equation}
whereas the quadratic metric (also referred to as rms metric) is given as
\begin{equation}
{\rm dev}_2 \ = \ \sqrt{ {1 \over N} \sum_{i=1}^N {\vert F_{\rm LTW}^{\ast} - F_{\rm LTW} \vert}^2 } \ .
\end{equation}
Here $F_{\rm LTW}$ refers to the wave energy flux of the detailed models, whereas
$F_{\rm LTW}^{\ast}$ refers to the wave energy flux obtained by the fitting formula
(or vice versa).  $N$ denotes the number of stars per test series. 

The test results have been deduced in a separate manner for the different values of
$\eta$ and $\alpha$; see Table~A.1 for details.  It is found that the average deviation,
regardless of the selected metric, are typically considerably better than 10\%,
and for some of the test series, the average deviation is found to be better
than 5\%.  We also checked the maximal deviation for individual stars, denoted as
$\Delta_{\rm max}$, for a given test series.  Our results indicate that the maximal
deviation never exceeds 15\%.
Finally, we also calculated the mean deviation between the model data and the
data given by the formula for the entire set of theoretical main-sequence stars,
comprised of 63 models.  We found that for the entire set of model stars,
the linear metric yields a mean deviation of 5.5\%, whereas the quadratic metric
yields a mean deviation of 6.5\%, a strong testimony of the quality of our fitting
formula for the overall range of solar-type stars.

\begin{table}
\caption{Accuracy of the fitting formula}
\centering
\vspace{0.05in}
\vspace{0.05in}
\begin{tabular}{l c c c c}
\hline
\hline
\noalign{\vspace{0.03in}}
$\alpha$ & Metric$^a$ & $\eta = 0.75$ & $\eta = 0.85$ & $\eta = 0.95$ \\
\noalign{\vspace{0.03in}}
\hline
\noalign{\vspace{0.03in}}
 1.5  & dev$_1$             &  5.2  &  6.1 &  5.6  \\
 1.5  & dev$_2$             &  5.8  &  7.5 &  6.2  \\
 1.5  & $\Delta_{\rm max}$  &  9.5  & 15.0 &  9.1  \\
\noalign{\vspace{0.03in}}
\hline
\noalign{\vspace{0.03in}}
 1.8  & dev$_1$             &   6.1 &  3.4 &  2.9  \\
 1.8  & dev$_2$             &   7.1 &  4.0 &  3.6  \\
 1.8  & $\Delta_{\rm max}$  &  11.7 &  6.5 &  6.5  \\
\noalign{\vspace{0.03in}}
\hline
\noalign{\vspace{0.03in}}
 2.0  & dev$_1$             &   6.9 &  7.3 &  5.8  \\
 2.0  & dev$_2$             &   7.3 &  7.9 &  7.3  \\
 2.0  & $\Delta_{\rm max}$  &  12.4 & 10.1 & 15.5  \\
\noalign{\vspace{0.03in}}
\hline
\end{tabular}
\vspace{0.05in}
\begin{list}{}{}
\item[]$^a$Data for dev$_1$, dev$_2$ and $\Delta_{\rm max}$ are given in percent.
\end{list}
\end{table}

\bigskip

\begin{acknowledgements}

This work has been supported by the
Faculty of Engineering and Computer Sciences,
Izmir University of Economics (D.~E.~F.) and the
Department of Physics, University of Texas at Arlington (M.~C.).
The authors also appreciate previous comments by P. Ulmschneider
and Z.~E. Musielak.

\end{acknowledgements}

\bigskip

\end{document}